\documentclass{aa}

\usepackage{graphicx}
\usepackage{txfonts}
\usepackage{siunitx}
\DeclareSIUnit\gauss{G}
\usepackage{ragged2e}
\usepackage{natbib}
\usepackage{subfigure}
\usepackage{hyperref}
\hypersetup{
    colorlinks=true,
    linkcolor=blue,
    citecolor=blue,
    filecolor=magenta,      
    urlcolor=cyan,
}

\newcommand{\va}{v_{\rm A}}

\begin{document}

\title{Self-consistent equilibrium models of prominence thin threads heated by Alfvén waves propagating from the photosphere}

   \author{Lloren\c{c} Melis
          \inst{1,2}
          \and
          Roberto Soler\inst{1,2}
          \and
          Jaume Terradas\inst{1,2}}

   \institute{Departament de Física, Universitat de les Illes Balears, E-07122, Palma de Mallorca, Spain\\ \and Insitut d'Aplicacions Computacionals de Codi Comunitari (IAC3), Universitat de les Illes Balears, E-07122, Palma de Mallorca, Spain}

\titlerunning{Prominence thread equilibrium models including Alfv\'en wave heating}
 \authorrunning{Melis et al.}
   

\abstract{The fine structure of solar prominences is made by thin threads that outline the magnetic field lines. Observations show that transverse waves of Alfv\'enic  nature are ubiquitous in prominence threads. These waves are driven at the photosphere and propagate to prominences suspended in the corona. Heating due to Alfv\'en wave dissipation could be a relevant mechanism in the cool and partially ionised prominence plasma. Here, we explore the construction of 1D equilibrium models of prominence thin threads that satisfy energy balance between radiative losses, thermal conduction, and Alfv\'en wave heating. We assume the presence of a broadband driver  at the photosphere that launches Alfv\'en waves towards the prominence. An iterative method is implemented, in which the energy balance equation and the Alfv\'en wave equation are consecutively solved. From the energy balance equation and considering no wave heating initially, we compute the equilibrium profiles along the thread of the temperature, density, ionisation fraction, etc. On these equilibrium profiles, we use the Alfv\'en wave equation to compute the wave heating rate, which is then put back in the energy balance equation to obtain new equilibrium profiles, and so on. The process is repeated until convergence to a self-consistent thread model heated by Alfv\'en waves is achieved. We have obtained equilibrium models composed of a cold and dense thread, a extremely thin PCTR, and an extended coronal region. The length of the cold thread decreases with the temperature at the prominence core and increases with the Alfv\'en wave energy flux injected at the photosphere. However, equilibrium models are not possible for sufficiently large wave energy fluxes when the wave heating rate inside the cold thread becomes larger than radiative losses. The maximum value of the wave energy flux that allows an equilibrium depends on the prominence core temperature. This constrains the existence of thread equilibria in realistic conditions.}

   \maketitle
   
\section{Introduction}
Solar prominences  consist of masses of relatively cool and dense plasma suspended in the solar corona, whose physical properties are similar to those in the chromosphere \citep[see, e.g.,][]{vialengvold2015}. High-resolution observations have shown that prominences are composed by a myriad of thin  threads, which seem to outline particular field lines of the general magnetic structure of prominences  \citep[see, e.g.,][]{lin2011thread,martin2015prominences}. The mechanical equilibrium of prominences suspended above the photosphere can be established when the upward force provided by the magnetic field  compensates gravity \citep[see, e.g.,][]{parenti2014prominences,gilbert2015balance,heinzel2015prominences}. On the other hand, the energy balance in solar prominences is a problem not so well understood \citep{gilbert2015balance}. A detailed knowledge of the heating and cooling processes operating in prominences is needed to explain the temperatures at prominence cores, which are estimated to be in the range 7,000--9,000~K \citep[see][]{parenti2014prominences}. According to radiative-equilibrium models, heating due to radiative illumination is believed to be the main heating source \citep[e.g.,][]{heasley1976prominences,anzer1999balance,heinzel2010prominences,heinzel2012equilibrium}. Using a slab model, \citet{heinzel2012equilibrium} computed radiative-equilibrium temperatures within the expected range of values for a plasma composed of hydrogen alone. However, if CaII losses are added, much lower radiative-equilibrium temperatures are obtained and, for certain conditions, they can be as low as 4,400~K \citep[see details in][]{heinzel2012equilibrium}. To the best of our knowledge, the effect of additional coolants has not been explored, but it may be that even lower radiative-equilibrium temperatures would be obtained if additional coolant species are added. Therefore, although radiative heating would be dominant in prominences, we cannot presently discard that additional sources of heating are also playing a role.  The purpose of the present paper is to explore the role of Alfv\'en wave dissipation as a potential heating mechanism in solar prominences that could complement radiative heating.

Mechanical and thermal equilibrium models of prominence thin threads have been studied before. For instance, \citet{degenhardt1993} studied the equilibrium between cooling and an imposed heating without considering the effect of thermal conduction. More detailed equilibrium models of threads were constructed  by \citet{terradas2021thread}, who studied the balance between radiative cooling, thermal conduction, and  heating. However,  the heating considered by \citet{terradas2021thread} was arbitrarily imposed rather than being the result of an actual physical process. The present work follows the method of \citet{terradas2021thread} to construct equilibrium models but using a consistently computed heating from the dissipation of Alfv\'en waves.

Observations have shown that transverse waves are ubiquitous in fine structures of solar prominences \citep[see the review by][]{arregui2018}. These waves are interpreted as magnetohydrodynamic waves of Alfv\'enic nature \citep[see, e.g.,][]{lin2009,ballester2015waves}. There are strong indications that the transverse waves  observed in prominences are driven at the photosphere, where the prominence magnetic field is anchored \citep{hillier2013waves}. It has been shown that Alfv\'enic waves can travel from the photosphere to the coronal structures, like prominences or coronal loops, transporting a significant amount of energy \citep[see][]{soler2019transport,soler2021}. The existence of an efficient dissipation mechanism in the prominence plasma could provide a way to thermalise the wave energy. The fact that the prominence plasma is only partially ionised introduces important dissipation mechanisms for Alfv\'en waves, namely ambipolar diffusion due to ion-neutral collisions and enhanced Ohm's diffusion due to electron-neutral collisions \citep[see][]{ballester2018partial}. Thus, heating due to Alfv\'enic waves arises as a possible heating mechanism in solar prominences whose importance needs to be explored.

There are some previous works that have explored the role of Alfv\'en wave heating in prominences. Using monochromatic waves and considering ion-neutral collisions, \citet{pecseli2000} concluded that Alfvén wave heating may only compensate for a tiny part of radiative losses. However, these authors did not consider a more realistic broadband spectrum of waves, whose presence is observationally confirmed \citep{hillier2013waves}. Conversely, \citet{parenti2007} estimated that the heating produced by the dissipation of unresolved non-thermal motions in prominences, which were attributed to Alfv\'en waves, could indeed compensate for a large fraction of radiative losses. Later, \citet{soler2016}  used an idealised slab model with a straight  magnetic field transverse to the slab axis. The slab was filled with a homogeneous and partially ionised prominence plasma and was embedded in a fully ionised corona. No fine structure was considered.  Alfv\'en waves were launched towards the prominence slab  and heating  due to ion-neutral collisions was computed. \citet{soler2016} estimated that the heating  could account for about 10~$\%$ of radiative losses if a spectrum of wave periods between 0.1 to 100 s is considered. Although the model was simple, the results of \citet{soler2016} evidenced the potential of wave heating in solar prominences.

A further study of wave heating in solar prominences was done in \citet{melis2021thread}, where the model implemented was more elaborated. An important difference with the previous work by  \citet{soler2016}, is that \citet{melis2021thread} considered a model for a thin thread including longitudinal non-uniformity and the anchoring of the magnetic field at the ends of the model representing the base of the corona. In the central part of the thread, where the plasma is densest and coolest, partial ionization and the roles of Ohm's and ambipolar diffusion were included. In the model of \citet{melis2021thread}, Alfv\'en waves were driven at one end on the thread to mimic photospheric excitation. Their results showed that wave heating could compensate for radiative losses in the cool part of the thread, where the plasma was partially ionised, although the energy balance was not consistently satisfied because the density and temperature profiles were assumed ad hoc and the back reaction of wave heating upon those profiles was not considered.

The aim of the present work is to go further and continue the study of Alfv\'en wave heating in prominence thin threads. Here the goal is to investigate the effect of wave heating in the equilibrium of threads. To this end, we shall combine the method to compute the wave heating rate of \citet{melis2021thread} with the method to compute equilibrium models of \citet{terradas2021thread}. In the presence of wave heating, threads should tend to an equilibrium configuration where wave heating consistently affects the profiles of density, temperature, and other variables along the thread. To study such influence and to investigate under what conditions an equilibrium is possible, we will implement a self-consistent approach loosely inspired by that used in \citet{ofman1999} for heating by resonant absorption in coronal loops. Starting from a thread model that includes no wave heating, the wave heating rate will be computed following \citet{melis2021thread}. Then, following \citet{terradas2021thread}, the energy balance equation will be solved including wave heating, and a new thread model will be obtained. The wave heating rate will be re-calculated in the new thread model and energy balance will be imposed again. This, in turn, will lead  to a different thread model, and so on.  This scheme will be run iteratively until convergence to a self-consistent model is reached. 

This paper is structured as follows. Section \ref{sec:model} contains an explanation of the thread model, the basic equations, and the implementation of the self-consistent method. In Section \ref{sec:results}, we  present and discuss the results. Finally, some conclusions are given in Section \ref{sec:conclusions}.

\section{Method}
\label{sec:model}

\subsection{Thread models satisfying energy balance}

\begin{figure*}
    \centering
    \includegraphics[width=2\columnwidth]{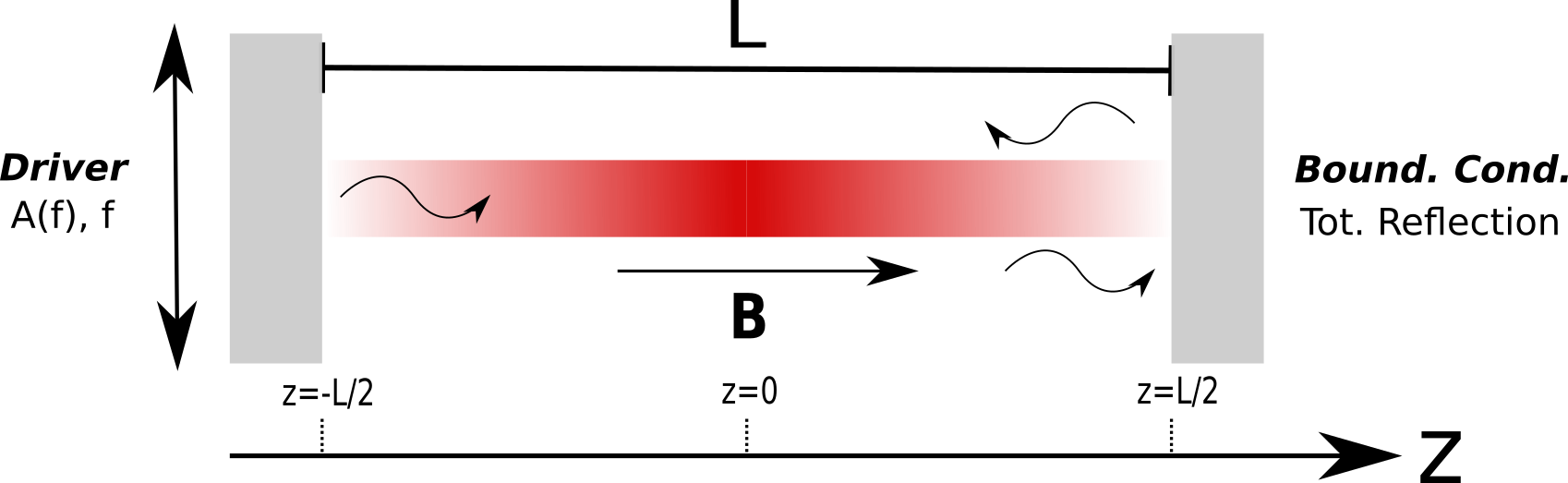}
    \caption{Sketch of the 1D thread model. The grey areas denote the photosphere, where the thread ends are anchored. The red color gradient aims to represent the density distribution along the thread.}
    \label{fig:model}
\end{figure*}

As in \citet{melis2021thread}, we use a 1D prominence thread model composed of a single magnetic field line of length $L=10^{8}$~m. Curvature and gravity are not considered, hence the magnetic field line is straight with a uniform value of $B=10$~G. We align the magnetic field with the $z$-axis for convenience. The centre of the thread is located at $z=0$, whereas its ends are located at $z=\pm L/2$. For simplicity, we assume a uniform gas pressure of $p=5 \times 10^{-3}$~Pa. A sketch of the thread model is shown in Figure \ref{fig:model}.

The density, $\rho$, and temperature, $T$, profiles are expected to be non-uniform along the thread, but $p$ is assumed to be uniform in order to preserve the hydrostatic equilibrium. Density and temperature are related with the pressure through the ideal gas law, which for a hydrogen plasma is
\begin{equation}
    p=(1+\xi_{i})\rho R T,
    \label{eq:gas}
\end{equation}
where $\xi_{i}$ is the ionisation fraction defined as the ratio between ion density with total density and $R$ is the ideal gas constant. In the cool parts of the thread, we expect the plasma to be partially ionised and so $\xi_i < 1$. In order to realistically obtain the ionisation fraction in a solar prominence thread, we would need to solve the full radiative transfer equations including incident radiation. This is far beyond the aims of this work. Instead, we approximate the ionization degree using the tabulated values given in \citet{heinzel2015transfer}, which are based on non-LTE computations for 1D prominence slabs. In Table~\ref{tab:heinzel} we indicate the values of the ionisation fraction as a function of the temperature for the pressure value considered here. The data of Table~\ref{tab:heinzel} is interpolated for intermediate temperatures and $\xi_{i}=1$ is imposed for $T \geq$~20,000~K. If either $T$  or $\rho$ is known, Equation~(\ref{eq:gas}) can be used to compute the other variable. Hence, we shall focus on the computation of $T$.

\begin{table}
    \centering
    \begin{tabular}{|c|c|c|c|c|c|}
    \hline
     \textbf{$T$} & 6000 & 8000 & 10000 & 12000 & 14000 \\ \hline
     \textbf{$\xi_{i}$} & 0.41 & 0.52 & 0.68 & 0.81 & 0.89 \\ \hline 
    \end{tabular}
    \caption{Values of the ionization fraction, $\xi_{i}$, for various values of the temperature (in K) for a gas pressure of $5 \times 10^{-3}$~Pa and an altitude above the photosphere of 20,000~km. Adapted from \citet{heinzel2015transfer}}
    \label{tab:heinzel}
\end{table}

In order to obtain the temperature profile, we  solve the energy balance equation  given by
\begin{equation}
\langle Q \rangle -\nabla \cdot \Vec{q} - \mathcal{L} + C = 0,
    \label{eq:balance}
\end{equation}
where  $\langle Q \rangle$ denotes the heating rate (this will be the wave heating, as explained later), $\Vec{q}=-\kappa_\parallel \nabla T$ is the heat flux due to thermal conduction with $\kappa_\parallel$  the parallel thermal conductivity, $\mathcal{L}$ represents the energy losses caused by radiative cooling, and $C$ may represent other heating sources like, e.g., heating due to viscous dissipation or radiative heating. Alfv\'en wave dissipation due to viscosity is less efficient than ambipolar dissipation in prominences, while a consistent treatment of radiative heating is well beyond our aims.  Therefore, these additional sources are ignored here.

In a partially ionised plasma, the parallel component of the thermal conductivity with respect to the magnetic field can be approximated by $\kappa_\parallel \approx \kappa_{e} + \kappa_{n}$, where $\kappa_{e}$ and $\kappa_{n}$ are the contributions of electrons and neutrals, respectively, which are given by
\begin{eqnarray}
    \kappa_{e}&=&3.2\frac{n_{e}^{2} k_{B}^{2} T}{\alpha_{e}+\alpha_{ee}}, \\
    \kappa_{n}&=&\frac{5}{3}\frac{n_{n}^{2} k_{B}^{2} T}{\alpha_{n}+\alpha_{nn}},
\end{eqnarray}
where $k_{B}$ is the Boltzmann constant, $n_{e}$ and $n_{n}$ are the number density of electrons and neutrals, respectively, $\alpha_{e}$ and $\alpha_{n}$ are the corresponding electron and neutral total friction coefficients, and $\alpha_{ee}$ and $\alpha_{nn}$ are the electron and neutral friction coefficients accounting for self-collisions.  The total  friction coefficients are computed as
\begin{equation}
    \alpha_{\beta}= \sum_{\beta \neq \beta'} \alpha_{\beta \beta'},
\end{equation}
where $\beta$ and $\beta'$ denote the different species that collide, that could be electrons (e), neutrals (n) or ions (i), and $\alpha_{\beta \beta'}=\alpha_{\beta' \beta}$ are the symmetric friction coefficients between individual particles. If the collision is between charged particles, the friction coefficient is
\begin{equation}
    \alpha_{\beta \beta'}=\frac{n_{\beta} n_{\beta'}e^{4} \ln{\Lambda_{\beta \beta'}}}{6 \pi \sqrt{2 \pi}\epsilon_{0}^2 m_{\beta \beta'}\left( k_{B} T/m_{\beta \beta'}\right)^{3/2}},
\end{equation}
where $m_{\beta \beta'}= m_{\beta}m_{\beta'}/\left( m_{\beta}+ m_{\beta'}\right)$ is the reduced mass,
$e$ is the electron charge,  $\epsilon_{0}$ is the electrical permittivity, $n_{\beta}$ is the number density, and $\ln\Lambda_{\beta \beta'}$ is the Coulomb's logarithm, expressed as
\begin{equation}
    \ln{\Lambda_{\beta \beta'}} = \ln{\left(\frac{24\pi \epsilon_{0}^{3/2}k_{B}^{3/2}T^{3/2}}{e^{3}\sqrt{n_{\beta}+n_{\beta'}}}\right)}.
\end{equation}
If at least one particle is neutral, the friction coefficient is 
\begin{equation}
    \alpha_{\beta n}= n_{\beta} n_{n} m_{\beta n} \left[ \frac{8 k_{B} T}{\pi m_{\beta n}}  \right]^{1/2} \sigma_{\beta n},
\end{equation}
where $\sigma_{\beta n}$ is the collision cross-section. See details in \citet{melis2021thread}.

The radiative cooling rate is computed using the function provided in \citet{athay1986radiation}, which is frequently used in the literature to approximate radiative losses in cool prominence plasmas \citep[see, e.g.,][]{mok1990}. We use Athay's function because it provides lower radiative losses than other cooling tables based on the optically thin approximation that may overestimate the actual losses for cool prominence temperatures. However, we note that other approaches are possible \citep[see][]{Dalgarno1972}, which more recent cooling functions implement \citep[e.g.,][]{Schure2009,Hermans2021,Brughmans2022}. The adopted radiative function is written down as
\begin{equation}
    \mathcal{L}(\rho,T)=f_{p}(T)\frac{\rho^{2}T^{2}}{m_p^{2}},
    \label{eq:radiative}
\end{equation}
where $m_{p}$ is the proton mass and $f_{p}$ is an analytical function of the temperature that  can be cast in MKS units as
\begin{eqnarray}
f_{p}(T) &=& 10^{-35} T^{-2} \Bigl\{ 0.4 \exp \left[ -30 \left( \log_{10} T -4.6 \right)^{2} \right] \nonumber \\ 
&&+ 4 \exp \left[ -20 \left( \log_{10} T -4.9 \right)^{2} \right] \nonumber \\ 
&&+ 4.5 \exp \left[ -16 \left( \log_{10} T - 5.35 \right)^{2} \right] \nonumber \\
&&+ 2 \exp \left[ -4 \left( \log_{10} T -6.1 \right)^{2} \right] \Bigr\}.
\label{eq:athay}
\end{eqnarray}

We note that because of the variation along the thread of the temperature, density, and ionization degree, both the thermal conductivity, $\kappa_\parallel$, and the radiative cooling rate, $\mathcal{L}$, are functions of $z$. For a given profile of the heating rate,  $\langle Q \rangle$, Equation~(\ref{eq:balance}) results in a complicated second order ordinary differential equation for the temperature profile, $T$, that needs to be solved numerically. To do so, we consider the same boundary conditions as in \citet{terradas2021thread}. We prescribe the temperature at the thread centre, $T_0$, and impose that this must correspond to the temperature minimum, so that  the boundary conditions are 
\begin{eqnarray}
T = T_{0}, \qquad \textrm{at} \qquad z=0,\\ \label{eq:central}
\frac{\partial T}{\partial z} = 0, \qquad \textrm{at} \qquad z=0. \label{eq:symmetry}
\end{eqnarray}
\citet{terradas2021thread} showed that the following relation must be satisfied at the thread centre, 
\begin{equation}
    \frac{\partial^2 T}{\partial z^2}= \frac{\mathcal{L}-\langle Q \rangle}{\kappa_{\parallel}} > 0, \qquad \textrm{at} \qquad z=0.
    \label{eq:second}
\end{equation}
For a cold thread, the temperature at the centre must be a minimum. According to Equation~(\ref{eq:second}), this is satisfied only when $\mathcal{L}> \langle Q \rangle$ at $z=0$. This gives us a restriction on the value of the heating rate that will be explored further.

The numerical integration of Equation~(\ref{eq:balance}) is performed  in two separate stages,  first from $z=0$ to $z=L/2$ and later from $z=0$ to $z=-L/2$. Then, the two solutions are joined together to construct the whole profile. This is done to correctly account for  asymmetries in the heating rate, $\langle Q \rangle$. The integration is performed in \emph{Wolfram Mathematica} with the routine \verb|NDSolve|, which automatically adapts the step size to minimise numerical errors.

\subsection{Alfv\'en wave heating}
\label{sec:waveheating}

As in \citet{melis2021thread}, the heating rate $\langle Q \rangle$ is computed from the dissipation of Alfv\'en waves. In the single-fluid magnetohydrodynamic approximation and assuming  linear waves, an equation that governs the stationary-state propagation of Alfv\'en waves along the thread was derived by \citet{melis2021thread}, namely
\begin{equation}
    \frac{\partial^{2} B_{\perp}}{\partial z^{2}} + \frac{\frac{\partial}{\partial z} \left( v_{\rm A}^{2} - i\omega \eta_{\rm C} \right)}{\left( v_{\rm A}^{2} - i\omega \eta_{\rm C} \right)}\frac{\partial B_{\perp}}{\partial z} +\frac{\omega^{2}}{\left( v_{\rm A}^{2} - i\omega \eta_{\rm C} \right)} B_{\perp}=0,
    \label{eq:basic}
\end{equation}
where $B_{\perp}$ is the transverse perturbation of the magnetic field, $\omega$ is the angular wave frequency, related with the linear frequency as $f = \omega/2\pi$, $v_{\rm A} = B/\sqrt{\mu_0\rho}$ is the Alfv\'en speed, with $\mu_0$ the magnetic permeability, and $\eta_{\rm C}$ is the Cowling's diffusion coefficient. Again, we note that both $v_{\rm A}$ and $\eta_{\rm C}$ are functions of $z$. 

Cowling's diffusion is the joint effect of Ohm's and ambipolar diffusion in a partially ionised plasma. Ohm's diffusion is caused by the collisions of electrons with other particles, whereas the ambipolar diffusion is related with the collisions between neutrals and charged particles. Their coefficients are written down as
\begin{eqnarray}
    \eta &=& \frac{\alpha_{e}}{\mu_{0}e^{2}n_{e}^{2}},\\
    \eta_{\rm A} &=& \frac{\xi_{n}^{2}}{\mu_{0}\alpha_{n}},
\end{eqnarray}
where $\eta$ is the Ohm's coefficient and $\eta_{\rm A}$ is the ambipolar coefficient, with $\xi_{n} = 1 - \xi_i$ the fraction of neutrals. Then, the Cowling's or total diffusion coefficient is computed as $\eta_{\rm C} = \eta + B^2 \eta_{\rm A}$.

The wave equation (Equation~(\ref{eq:basic})) needs to be solved considering appropriate boundary conditions at the ends of the thread. We assume that the wave driver in located at the left end of the thread, $z=-L/2$. Alfv\'en waves with a frequency, $f$, are driven  with a prescribed amplitude that depends upon the adopted  spectral weighting function, $A(f)$. Following \citet{tu2013heat} and \citet{arber2016chromo}, the spectral weighting function is assumed to be a piece-wise power-law function, expressed as
\begin{equation}
A\left(f\right) = A_0 \left\{ \begin{array}{lll}
\left(\frac{f}{f_b}\right)^{5/6}, & \textrm{if} & f \leq f_b, \\
\left(\frac{f}{f_b}\right)^{-5/6}, & \textrm{if} & f > f_b,
\end{array} \right. \label{eq:spectral}
\end{equation}
where $f_{b}$ is the break frequency and $A_{0}$ is a constant. The break frequency is set to 1.59 mHz as in \citet{tu2013heat}. The choice of this frequency is based on the observed spectrum of horizontal photospheric motions, which suggests that this frequency is between 1 and 10 mHz \citep[see, e.g.,][]{matsumoto2010spectra}. Physically, this break frequency should correspond to the beginning of the inertial range governed by the photospheric turbulence.  The constant $A_{0}$ depends on the  wave energy flux injected by the driver. The energy flux for an Alfv\'en wave of frequency $f$ averaged over a one full period $1/f$ is 
\begin{equation}
    \langle \vec{\Pi} \rangle = -\frac{1}{2 \mu_0} \textrm{Re} \left[ v_{\perp} B_{\perp}^{*} \right]\vec{B},
\end{equation}
where * denotes the complex conjugate and $v_\perp$ is the transverse velocity perturbation, which can be expressed in terms of $B_\perp$ as \citep[see][]{melis2021thread},
\begin{equation}
    v_\perp = \frac{i}{\omega}\frac{\va^2}{B}\frac{\partial B_{\perp}}{\partial z}.
\end{equation}
The energy flux can be decomposed as $\langle \vec{\Pi} \rangle$ = $\langle \vec{\Pi} \rangle^{\uparrow}$ - $\langle \vec{\Pi} \rangle^{\downarrow}$, where $\langle \vec{\Pi} \rangle^{\uparrow}$ and $\langle \vec{\Pi} \rangle^{\downarrow}$ are the parallel and anti-parallel components of $\langle \vec{\Pi} \rangle$ with respect to the direction of the background magnetic field. Their expressions are
\begin{eqnarray}
    \langle \vec{\Pi} \rangle^{\uparrow} &=& \frac{1}{8} \sqrt{\frac{\rho}{\mu_{0}}} Z^{\uparrow} Z^{\uparrow *} \vec{B},\\
    \langle \vec{\Pi} \rangle^{\downarrow} &=& \frac{1}{8} \sqrt{\frac{\rho}{\mu_{0}}} Z^{\downarrow} Z^{\downarrow *} \vec{B},
\end{eqnarray}
where $Z^{\uparrow,\downarrow} = v_{\perp} \mp \frac{1}{\sqrt{\mu_{0} \rho}}B_\perp$ are the so-called Els\"{a}sser variables that represent parallel-propagating and anti-parallel-propagating Alfv\'enic disturbances, respectively. At $z=-L/2$, the quantity $\langle \vec{\Pi} \rangle^{\uparrow}$ corresponds to the wave energy flux injected by the driver for one particular frequency, $f$. Thus,  $\Sigma_{f} \langle \boldsymbol{\Pi} \rangle^{\uparrow}$ is the total energy flux injected by the driver for all the frequencies in the spectrum. The constant $A_{0}$ is computed by assuming that the total injected energy flux is equal to a prescribed value, which is hereafter denoted  by $\langle \pi \rangle$.

The wave driver is actually located at the photosphere, but the thread model only includes the coronal part. It is known that the transmission of Alfv\'en waves from the photosphere to the corona is heavily influenced by the conditions in the chromosphere, where reflection and dissipation can  equally be important \citep[see][]{soler2017,soler2019transport}. To account for the chromospheric filtering, we use the empirical wave energy transmission coefficient from  \citet{soler2019transport}, $\mathcal{T}(f,B_{ph})$, which is a function of the wave frequency, $f$, and  the magnetic field strength at the photosphere, $B_{ph}$, namely \begin{equation}
\begin{split}
    \mathcal{T}(f,B_{ph}) \approx a_{0}(B_{ph}) \frac{1}{\sqrt{2\pi \sigma{(B_{ph})}^{2}}} \exp \left( -\frac{\left(  \log_{10} f -\mu(B_{ph})\right)^{2}}{2\sigma{(B_{ph})}^{2}} \right) \\
    \times \left[ 1 + \textrm{erf} \left( \frac{\alpha(B_{ph})}{\sqrt{2}} \frac{\log_{10}f - \mu(B_{ph})}{\sigma(B_{ph})} \right) \right],
\end{split}
\end{equation}
where erf is the error function, and $a_{0}$, $\mu$, $\sigma$ and $\alpha$ are the amplitude, location, scale and shape parameters, respectively, which are given in \citet{soler2019transport} and depend upon $B_{ph}$. Here, we take $B_{ph}=100$~G. Therefore, the effective spectral weighting function at the base of the corona is
\begin{equation}
    A_{\rm eff.}(f,B_{ph}) = A(f) \sqrt{\mathcal{T}(f,B_{ph})}.
\end{equation}
We note that $ A_{\rm eff.}$ is proportional to the square root of $\mathcal{T}(f,B_{ph})$ because there is quadratic relation between the wave energy and magnetic field perturbations.

\begin{figure}[!htbp]
    \resizebox{\hsize}{!}{\includegraphics{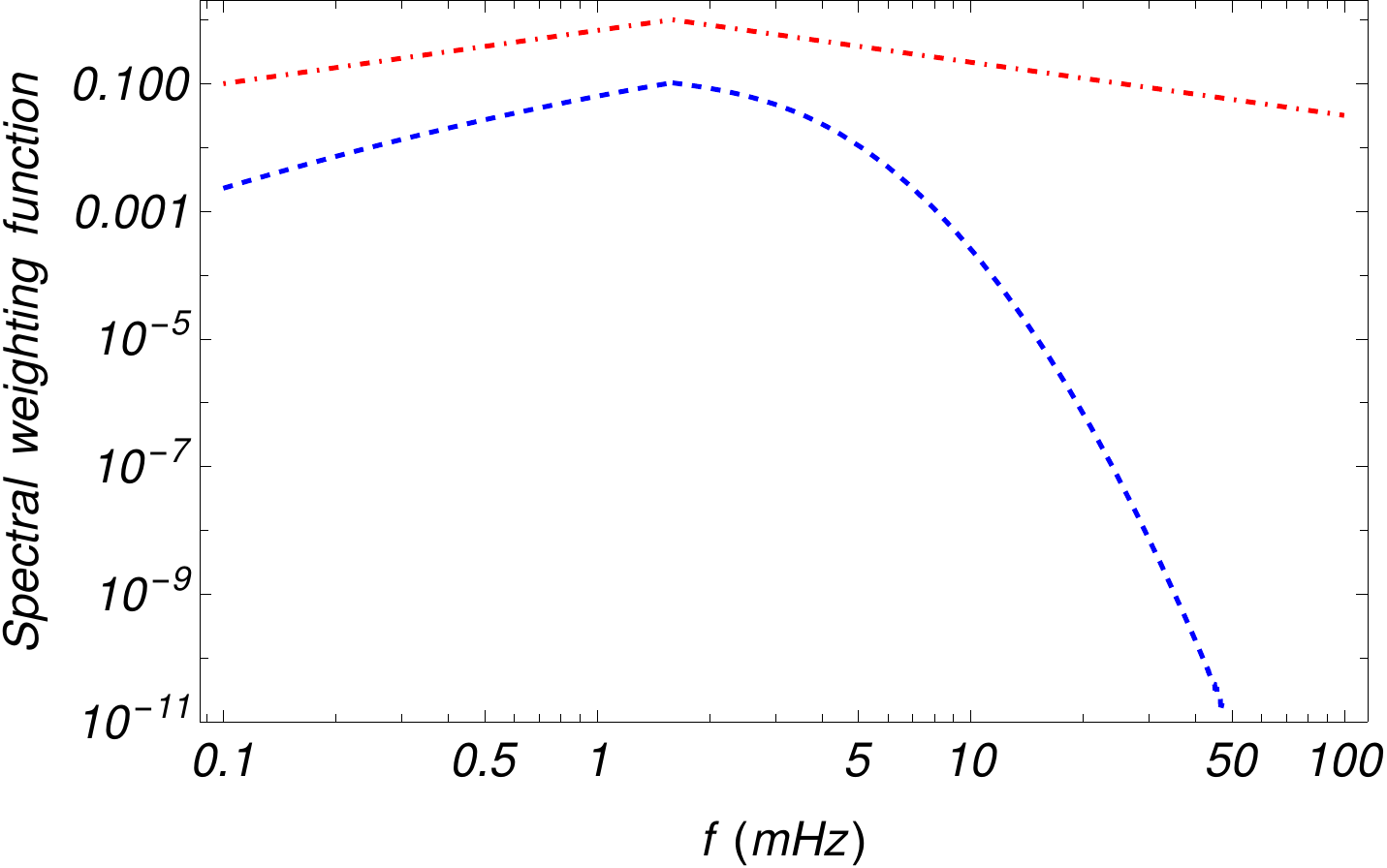}}
    \caption{Photospheric spectral weighting function (dash-dotted red line) and effective coronal spectral weighting function (dashed blue line) as functions of the wave driver frequency. Note that both axes are in logarithmic scale. For the purpose of this plot, we set $A_0=1$.}
    \label{fig:spectral}
\end{figure}

Figure \ref{fig:spectral} compares the  photospheric spectral weighting function, $A(f)$, with the effective coronal one, $A_{\rm eff.}(f) $. Reflection at the chromosphere decreases the spectrum amplitude  in the low frequency range, while the strong ion-neutral damping produces the rapid decrease of the amplitude as the frequency increases. Indeed, the highest frequencies in the spectrum are effectively suppressed by the chromospheric filtering.

Hence, the form of the magnetic field perturbation at the driver location is prescribed as
\begin{equation}
    B_{\perp} = A_{\rm eff.}(f,B_{ph})  \exp \left( i \Phi(f) \right) \qquad \textrm{at} \qquad z= - \frac{L}{2},
    \label{eq:magnetic}
\end{equation}
where  $0<\Phi(f)<2\pi$ represents random phase different for each frequency in the spectrum. On the other hand, the right end of the thread is treated as a perfectly reflecting boundary for the waves. Contrary to the empirical transmissivity used at the left boundary, there is no available empirical reflection coefficient to the used at the right boundary to account for the effect of the chromosphere, so that  a boundary condition has to be imposed arbitrarily. \citet{melis2021thread} tested different boundary conditions, namely total reflection, total transmission, and partial reflection. Concerning the wave heating rate in the cool part of the thread, they found no significant differences between the results for the different boundary conditions. The physical explanation for this finding is that the high-frequency waves that actually produce most of the heating have short damping lengths and are almost completely dissipated during their first passing through the thread, so that little energy is left to reach and reflect at the boundary. Thus, we restrict ourselves to the perfectly reflecting scenario, for which the boundary condition at the right end is 
\begin{equation}
    \frac{\partial B_{\perp}}{\partial z} = 0 \qquad \textrm{at} \qquad z = \frac{L}{2}. \label{eq:boundary}
\end{equation}

We discretise the spectrum of the driver into  101 individual  frequencies between 0.1 and 100~mHz with a logarithmic spacing. For each one of the frequencies in the discretised spectrum, Equation~(\ref{eq:basic}) with the above boundary conditions (Equations~(\ref{eq:magnetic}) and (\ref{eq:boundary})) is numerically solved in \emph{Wolfram Mathematica} using again the routine \verb|NDSolve|. However, now we use finite elements to perform the integration with a customised non-uniform mesh whose resolution depends upon the considered frequency, $f$. Results from \citet{terradas2021thread} point out that temperature and density profiles are expected to have sharp variations at the prominence-corona transition region (PCTR). Concerning the Alfv\'en waves, the PCTR needs to be resolved with a sufficiently high resolution to correctly compute the transmission of the waves into the dense part of the thread. Otherwise, artificial reflections may happen, which can heavily influence the results. The custom grid is divided into five different regions: the central part where the plasma is coolest and partially ionised, two PCTR zones surrounding the centre where there is the sharp variation of temperature and density, and two outermost parts that represent the coronal regions. For a prescribed wave frequency, the mesh resolution in the central and coronal regions is determined by the local Alfv\'en wavelength at the centre, $\lambda_{0} = \va(z=0)/f$, or at the ends, $\lambda_{c}=\va(z=L/2)/f$, of the thread, respectively. We set the resolution in each region equal to a small fraction of their respective local wavelengths, being $\lambda_{0}/100$ for the central region and $\lambda_{c}/20$ for the coronal regions. In the PCTR zones, a fixed mesh resolution of 100~m is used for all frequencies. The width of the two high-resolution PCTR zones is set to 40~km and their location is determined by a routine that detects the position at which the derivative of the temperature profile with respect to $z$ is maximum or minimum.

The numerical solution of Equation~(\ref{eq:basic}) provides us with the magnetic field perturbation, $B_\perp$, along the thread caused by the propagation of Alfv\'en waves with frequency, $f$.  The next step is to compute the wave heating rate. The plasma heating  is the consequence of the wave energy absorption, i.e., the part of the wave energy flux that is deposited into the plasma due to Cowling's diffusion. The instantaneous heating rate is computed as
\begin{equation}
    Q_f = \mu_{0} \eta_{\rm C} \left| \vec{j_{\perp}} \right|^2,
    \label{eq:heat}
\end{equation}
where  $\vec{j_{\perp}}$ is the perpendicular component of the current density given by
\begin{equation}
    \vec{j_{\perp}} = \frac{1}{\mu_{0}} \frac{\partial B_{\perp}}{\partial z} \hat{e}_{\perp},
\end{equation}
where $\hat{e}_{\perp}$ denotes the unit vector in the direction perpendicular to the magnetic field. The instantaneous heating rate needs to be averaged over one full period of the wave, $1/f$, to obtain the net heating produced by that particular frequency, namely
\begin{equation}
    \langle Q \rangle_{f} = \frac{\eta_{\rm C}}{2 \mu_{0}} \left| \frac{\partial B_{\perp}}{\partial z} \right|^{2}.
    \label{eq:heating}
\end{equation}
In order to obtain the total heating rate, we add together the net heatings produced by all the frequencies in the broadband spectrum. Thus,
\begin{equation}
     \langle Q \rangle = \sum_f \langle Q \rangle_{f}.
\end{equation}
The heating profile computed in such a way is the one that is used in the energy balance equation (Equation~(\ref{eq:balance})).

\subsection{Self-consistent strategy}

The goal is to obtain prominence thread models in which the Alfv\'en wave heating obtained from the solution of Equation~(\ref{eq:basic}) satisfies the energy balance condition (Equation~(\ref{eq:balance})). Thus, threads heated by Alfv\'en waves would be in thermal equilibrium. This equilibrium is studied in terms of two parameters: the central temperature, $T_0$, and the wave energy flux injected at the photosphere, $\langle \pi \rangle$. For $T_{0}$, we consider values between 7,000 to 10,000 K, as consistent with the  temperatures in prominence cores. Regarding the injected energy flux, it is expected that the wave heating rate increases with the injected flux. Thus,  we will progressively increase the value of $\langle \pi \rangle$ until the requirement of Equation~(\ref{eq:second}) is no longer satisfied and so no equilibrium models are possible for such conditions.

The construction of self-consistent thread models heated by Alfv\'en waves poses a circular problem. We must solve Equation~(\ref{eq:basic}) to compute the wave heating rate. Since  the Alfv\'en speed and the Cowling's coefficient appear in Equation~(\ref{eq:basic}), we need the temperature and density profiles to be known beforehand. However, the temperature profile is computed from  Equation~(\ref{eq:balance}), which in turn require the wave heating rate to be known. The solution to this circular problem is here attacked using an iterative strategy.

\begin{figure}[!htbp]
    \resizebox{\hsize}{!}{\includegraphics{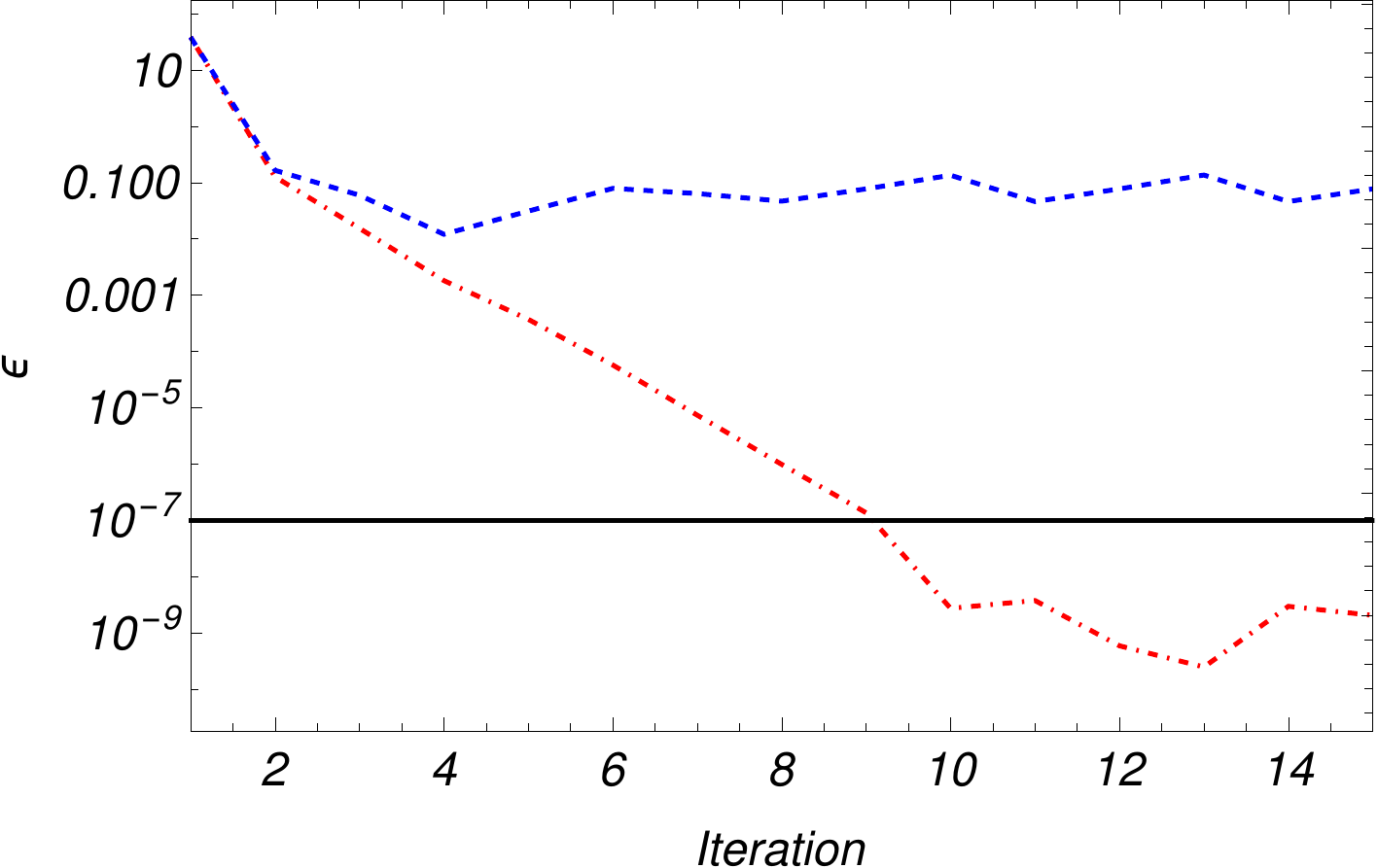}}
    \caption{Variation of the parameter $\varepsilon$ with the number of iterations for a case that converges (dash-dotted red line) and a case that does not converge (dashed blue line). The tolerance value of $\varepsilon_0 = 10^{-7}$ is denoted by a horizontal line. Note that the vertical axis is in logarithmic scale.}
    \label{fig:eps}
\end{figure}

To start with, we select particular values of $T_0$ and $\langle \pi \rangle$ for which the self-consistent thread model is to be computed. A thread model satisfying the energy balance condition (Equation~(\ref{eq:balance})) is obtained for the case with no wave heating, i.e., $ \langle Q \rangle = 0$. Subsequently, the Alfv\'en wave equation (Equation~(\ref{eq:basic})) is solved in that initial thread model and the wave heating rate is obtained. This heating rate is put back in  Equation~(\ref{eq:balance}) and a new thread model, now including wave heating, is computed. This completes the first iteration of the process. Generally, the initial thread model obtained for $ \langle Q \rangle = 0$ and that obtained after the first iteration would be different. To quantify the difference between models we study the variation of the  length of the cool part of the thread, hereafter called the `thread length' and denoted by $a$.  The thread lengths are computed by automatically detecting the position of the two PCTR zones, as explained in Section~\ref{sec:waveheating}. The location of PCTR and so the thread length are computed by using the second derivative of temperature and detecting the values of $z$ for which it is zero. We use the parameter $\varepsilon$, which is defined as the relative error norm of the thread length, namely
\begin{equation}
\varepsilon = \frac{|a_{i}-a_{i-1}|}{a_{i}},
\end{equation}
where $a_{i}$ and $a_{i-1}$ are the thread lengths for the iteration $i$ and the previous iteration, respectively. We shall assume that a self-consistent model has been achieved when $\varepsilon < \varepsilon_0$, where $\varepsilon_0$ is a prescribed small tolerance. We use $\varepsilon_0 = 10^{-7}$. Therefore, the process explained before for the first iteration would be repeated as many times as necessary until converge, if possible, is reached. If the requirement of Equation~(\ref{eq:second}) is satisfied, the value of $\varepsilon$ decreases with the number of iterations, leading to successful convergence. However, if the requirement of Equation~(\ref{eq:second}) is not satisfied, convergence can never be achieved and the value of $\varepsilon$ may even increase with the number of iterations. Figure~\ref{fig:eps} displays the variation of $\varepsilon$ with the number of iterations in two example cases: a case in which the chosen values of $T_0$ and $\langle \pi \rangle$ allow convergence and a self-consistent model is obtained and another case that does not converge.

\begin{figure*}
    \centering
    \includegraphics[width=2\columnwidth]{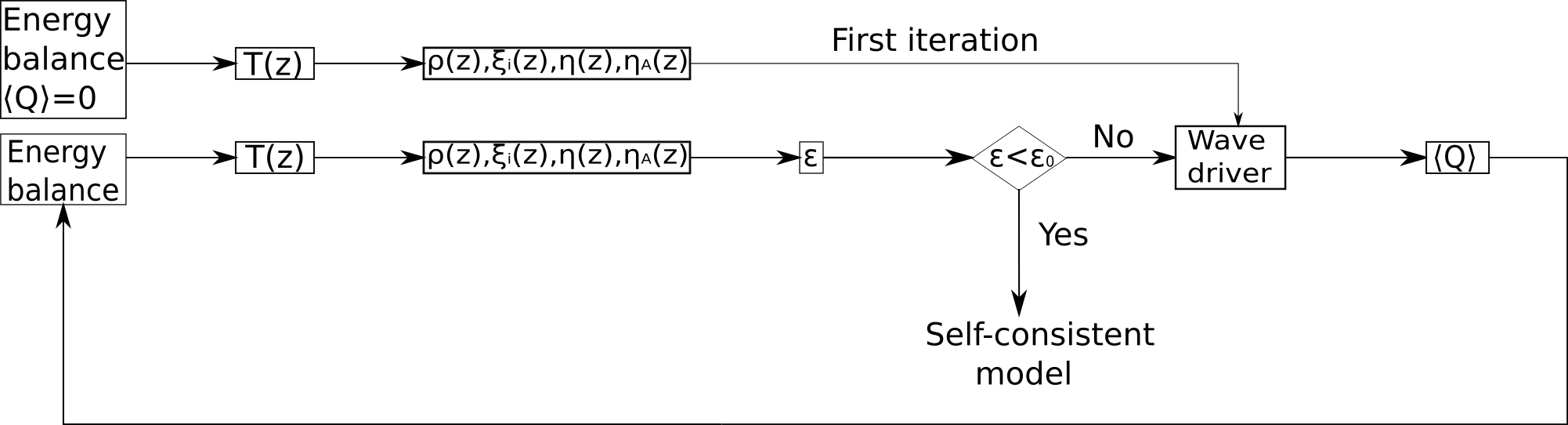}
    \caption{Flux diagram of the iterative strategy that we use to compute self-consistent thread models. All parameters are defined in the text.}
    \label{fig:diagram}
\end{figure*}

Figure~\ref{fig:diagram} summarises in a schematic way how the iterative strategy works to compute a self-consistent model. 

\section{Results}
\label{sec:results}


\subsection{A typical thread model}

We start the presentation of results by analysing a typical thread model obtained for a certain combination of $T_0$ and $\langle \pi \rangle$. This will help us to understand better the results of the subsequent parameter study.

\begin{figure*}
    \centering
    \includegraphics[width=2\columnwidth]{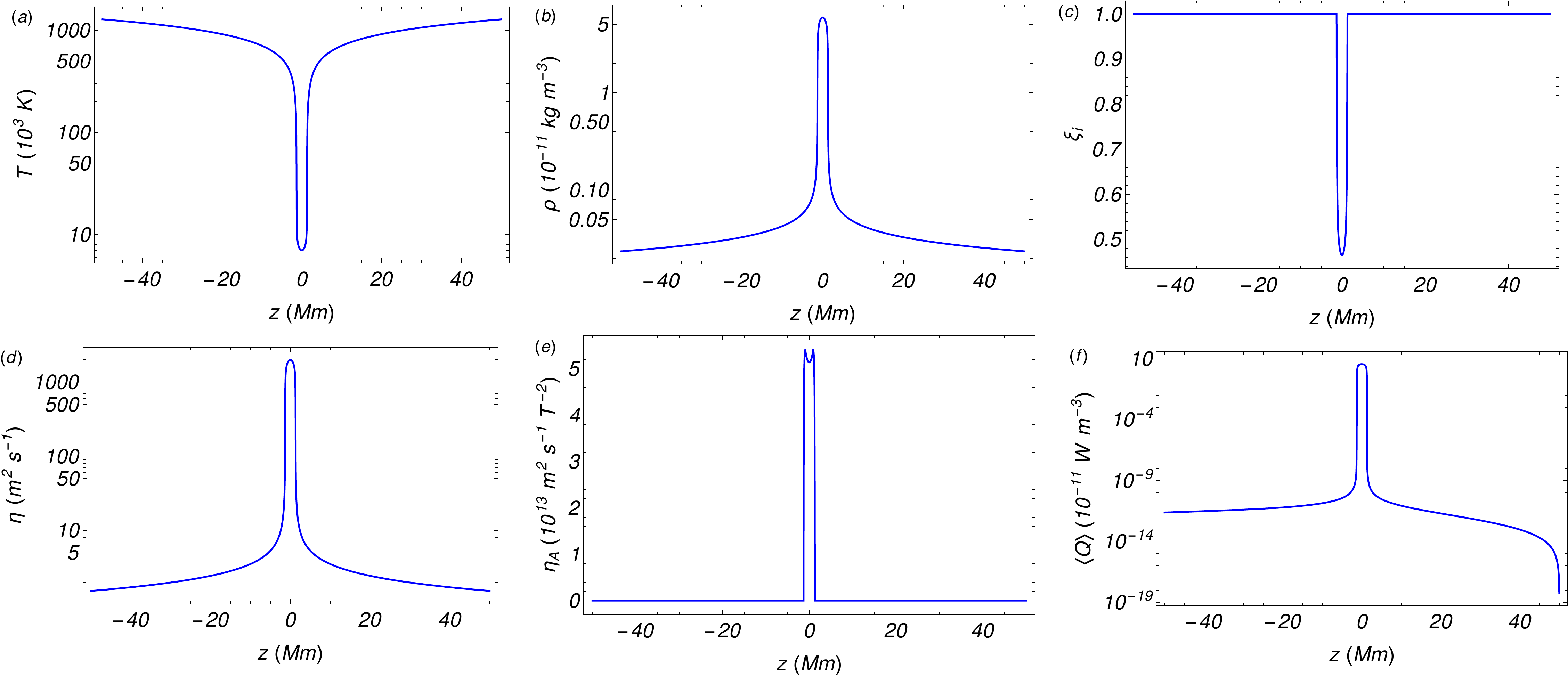}
    \caption{Equilibrium profiles along the thread: (\textit{a})  temperature, (\textit{b}) density, (\textit{c}) ionisation fraction, (\textit{d}) Ohm's diffusion coefficient, (\textit{e}) ambipolar diffusion coefficient, and (\textit{f}) volumetric wave heating rate. Results for $T_0 =$~7,000~K and $\langle \pi \rangle =$~0.2~W~m$^{-2}$.}
    \label{fig:back_full}
\end{figure*}

Hence, we consider $T_0 =$~7,000~K and $\langle \pi \rangle =$~0.2~W~m$^{-2}$. Figure \ref{fig:back_full} displays the profiles of temperature, density, ionization fraction, Ohm's and ambipolar coefficients, and heating rate along the thread. The temperature profile (Figure~\ref{fig:back_full}a) is directly computed from Equation~(\ref{eq:balance}) and its shape is similar to that obtained in the models of \citet{terradas2021thread} with a constant heating term. The temperature profile can be divided into three distinct regions: the cold central region  with a parabolic-like shape around the temperature minimum, then a sharp increase of the temperature by several orders of magnitude in a very thin layer that clearly matches the expected PCTR mentioned before, and, finally, a wider region with a more gentle increase of temperature that extends up to the ends of the thread, where typical coronal temperatures of order of $\sim 10^{6}$ K are reached. The temperature profile is largely symmetric with respect to $z=0$. Very small asymmetries exist, which are not noticeable at the scale of the figure. In this particular model, the thread length is $a\approx$~2.66~Mm, which corresponds to a small fraction of the total length of the magnetic field line, namely $a/L \approx 0.027$.

Figure \ref{fig:back_full}b shows the density profile, which is computed from the temperature profile with the help of Equation~(\ref{eq:gas}). The density adopts an inverse profile with respect to that of the temperature. The largest density of $\sim 10^{-11}$~kg~m$^{-3}$ is located at the centre. In the outermost coronal part of the model, the density decreases to $\sim 10^{-13}$~kg~m$^{-3}$ near the ends. The ionisation fraction is plotted in Figure \ref{fig:back_full}c. We recall that the ionisation fraction is  computed from the temperature using Table~\ref{tab:heinzel}. We obtain a very narrow partially ionised region around the thread centre, where the plasma is coldest and densest, with a minimum of $\xi_i \approx 0.5$. The plasma rapidly gets fully ionised in the PCTR, so that  $\xi_i = 1$ in the coronal part of the model.

The Ohm's diffusion coefficient, $\eta$, can be seen in Figure \ref{fig:back_full}d. Essentially, the spatial profile of $\eta$ follows that of the density. The maximum of $\eta$ is located at the central part, $\eta \sim$~\SI{1e3}{\metre\squared\per\second}, and decreases sharply towards the coronal part of the model, where $\eta \sim$~\SI{1}{\metre\squared\per\second}. Therefore, Ohm's diffusion is three orders of magnitude more efficient in the cool central part than in the hot coronal part. The ambipolar diffusion coefficient, $\eta_{\rm A}$, is shown in Figure \ref{fig:back_full}e. The ambipolar coefficient is zero in the coronal region of the model, where the plasma is fully ionised. Inside the partially ionised zone, $\eta_{\rm A}$ has two relative maxima slightly displaced from the centre of the thread. The maximum value is  $\eta_{\rm A}\sim$~\SI{1e13}{\metre\squared\per\second\per\tesla\squared}. Considering that the background magnetic field strength is $B=10$~G, the maximum value of the effective ambipolar diffusion coefficient is $B^2\eta_{\rm A}\sim$~\SI{1e7}{\metre\squared\per\second}, which points out that ambipolar diffusion is much more efficient than Ohm's diffusion in the partially ionised region.

The wave heating rate consistent with this particular model can be seen in Figure~\ref{fig:back_full}f. The heating is very non-uniform along the thread. When approaching the cold central region, the heating rate increases several orders of magnitude compared to the values obtained in the hot coronal part. Ambipolar diffusion is the dominant dissipation mechanism in our model and produces most of the heating in the cold partially ionised region. In addition, unlike the rest of quantities, the heating rate displays a clear asymmetry. Towards the right end, the heating rate tends to zero. The reason for this asymmetry resides in the fact that the wave driver is only located at the left end of the model, while the boundary condition at the right end imposes total reflection. However, the asymmetry is much less noticeable in the central region and, indeed, the profiles obtained for the rest of model quantities are nearly symmetric with respect to $z=0$.


\subsection{Effect of the central temperature}
\label{sec:temp}

\begin{figure*}
    \centering
    \includegraphics[width=2\columnwidth]{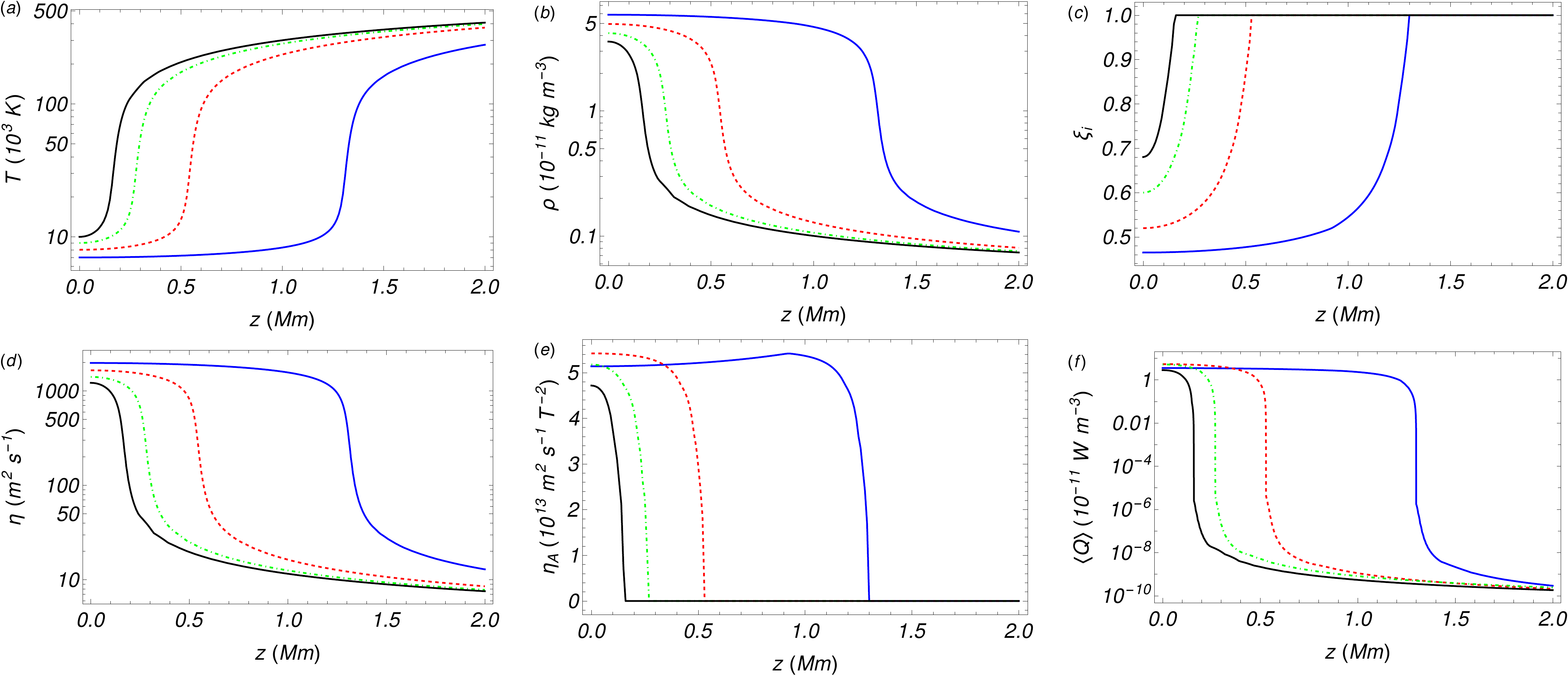}
    \caption{Same as Figure \ref{fig:back_full}, but for different values of the central temperature, namely $T_0=$~7,000~K (solid blue line), $T_0=$~8,000~K (dashed red line), $T_0=$~9,000~K (dash-dotted green line), and $T_0=$~10,000~K (solid black line). Only  a close-up view of the central part and the PCTR of the thread for $z \ge 0$ is displayed. Results for $\langle \pi \rangle =$~0.2~W~m$^{-2}$.}
    \label{fig:back_zoom}
\end{figure*}

After presenting the results for a particular case, here we explore the effect of changing the central temperature, $T_0$. In all cases, we keep $\langle \pi \rangle =$~0.2~W~m$^{-2}$ as before. According to \citet{terradas2021thread}, the thread length decreases when the central temperature increases. This result is confirmed in Figure \ref{fig:back_zoom}, which displays the equilibrium profiles for thread models with different values of the central temperature. Due to the near symmetry of the models, Figure \ref{fig:back_zoom} only shows the profiles for positive values of $z$ corresponding to the central part of the thread and the PCTR, because it is there where the main differences appear, while the profiles in the coronal part would be seen as nearly identical.

In agreement with \citet{terradas2021thread}, when the central temperature increases, the cold part of the thread becomes narrower. The decrease of the thread length can be explained by the relative importance of the various terms in the energy balance equation (Equation~(\ref{eq:balance})). If the effect of heating is small, which would happen  for the small wave energy flux assumed here, thermal conduction is mostly responsible for balancing radiative losses in the equilibrium \citep{terradas2021thread}. As $T_0$ increases, radiative losses become more and more efficient in the central region. Hence, thermal conduction requires a stronger temperature gradient to compensate for the increase of radiation, which results in narrower threads. 

Another consequence of the increase of the central temperature is that the  partially ionised plasma gets confined to a thinner region  and the lowest value of $\xi_i$ increases. This  is explained by the relation between temperature and ionisation fraction as  seen in Table~\ref{tab:heinzel}. The fact that the plasma gets more ionised as $T_0$ increases impacts on the efficiency of Ohm's and ambipolar diffusion. The coefficients associated with both effects become smaller in magnitude when $T_0$ increases.

Despite the fact that the thread length changes, the  shapes of the profiles for the different values of the central temperature are quite similar, except that for the ambipolar diffusion coefficient (Figure~\ref{fig:back_zoom}e). For high central temperatures, the maximum of $\eta_{\rm A}$ is located at the centre of the thread instead of being displaced from the centre when the central temperature is low. The ambipolar diffusion coefficient has a complicated dependence on the temperature and ionization degree that results in this peculiar behaviour. In turn, the heating rate (Figure~\ref{fig:back_zoom}f) has also a similar shape for all the central temperatures considered, but its value at the thread centre increases or decreases with $T_0$ according to the behaviour of the ambipolar diffusion coefficient at $z=0$.

\subsection{Effect of the wave energy flux}
\label{sec:flux}

\begin{figure*}
    \centering
    \includegraphics[width=2\columnwidth]{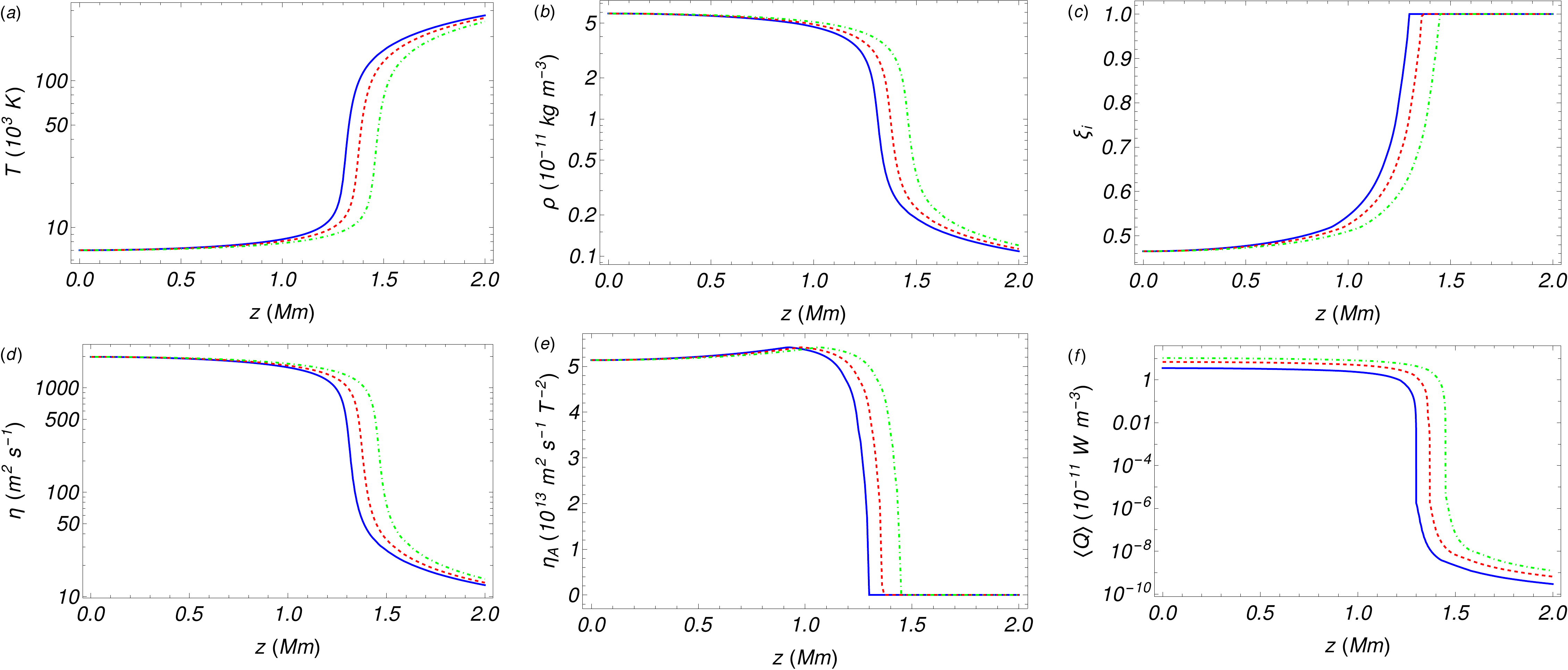}
    \caption{Same as Figure~\ref{fig:back_zoom} with $T_0 =7000$~K and three different values of injected wave energy flux, namely $\langle \pi \rangle =$~0.2~W~m$^{-2}$ (solid blue line), $\langle \pi \rangle =$~0.4~W~m$^{-2}$ (dashed red line), and $\langle \pi \rangle =$~0.8~W~m$^{-2}$ (dash-dotted green line).}
    \label{fig:back_flux}
\end{figure*}

Besides the role of the central temperature, which was already explored by \citet{terradas2021thread}, here we are more interested in the effect of the injected wave energy flux. To this end, we have set the central temperature to $T_0 =$~7,000~K and have considered three different values of the injected energy flux, namely $\langle \pi \rangle =$~0.2, 0.4, and 0.8~W~m$^{-2}$. The computed models are displayed in Figure~\ref{fig:back_flux} where, again, only positive values of $z$ corresponding to the central part of the thread and the PCTR are displayed. 

Two main effects of varying $\langle \pi \rangle$ are seen in Figure~\ref{fig:back_flux}. First of all, the larger $\langle \pi \rangle$, the larger the wave heating rate $\langle Q \rangle$ (see Figure~\ref{fig:back_flux}f). This result is not surprising, because the wave heating rate is expected to be proportional to the injected wave energy flux. In other words, the larger the available wave energy, the larger the energy that is dissipated. The second effect that is obvious in Figure~\ref{fig:back_flux} is the growth of the thread length as $\langle \pi \rangle$ increases. Indeed, this result is a consequence of the increase of the heating rate discussed before. Following \citet{terradas2021thread}, we can understand the increase of the thread length by using an approximate analytic expression for this quantity given by
\begin{equation}
    a \approx \sqrt{\frac{2 \kappa_\parallel T_0}{\mathcal{L}-\langle Q \rangle}}, \label{eq:aapprox}
\end{equation}
where $\kappa_\parallel$, $\mathcal{L}$, and $\langle Q \rangle$ need to be evaluated at the thread centre, $z=0$. If the central temperature remains constant and the wave heating rate increases at $z=0$ and approaches the minimum value of radiative loses, then the denominator in Equation~(\ref{eq:aapprox}) decreases. Consequently,  the thread length increases. In \citet{terradas2021thread} a similar conclusion is reached although for a spatially uniform heating term. Here, the heating rate is very non-uniform along the field line, but Equation~(\ref{eq:aapprox}) can still be used to understand to effect of the heating on the thread length. In this regard, and for a fixed central    temperature, only the value of $\langle Q \rangle$ at the centre seems to be relevant, while its spatial profile does not play a role according to Equation~(\ref{eq:aapprox}). The effect of the wave heating on the thread length is investigated further in Section~\ref{sec:param}.

\subsection{Role of wave heating in the energy balance}

\begin{figure*}
    \centering
    \includegraphics[width=2\columnwidth]{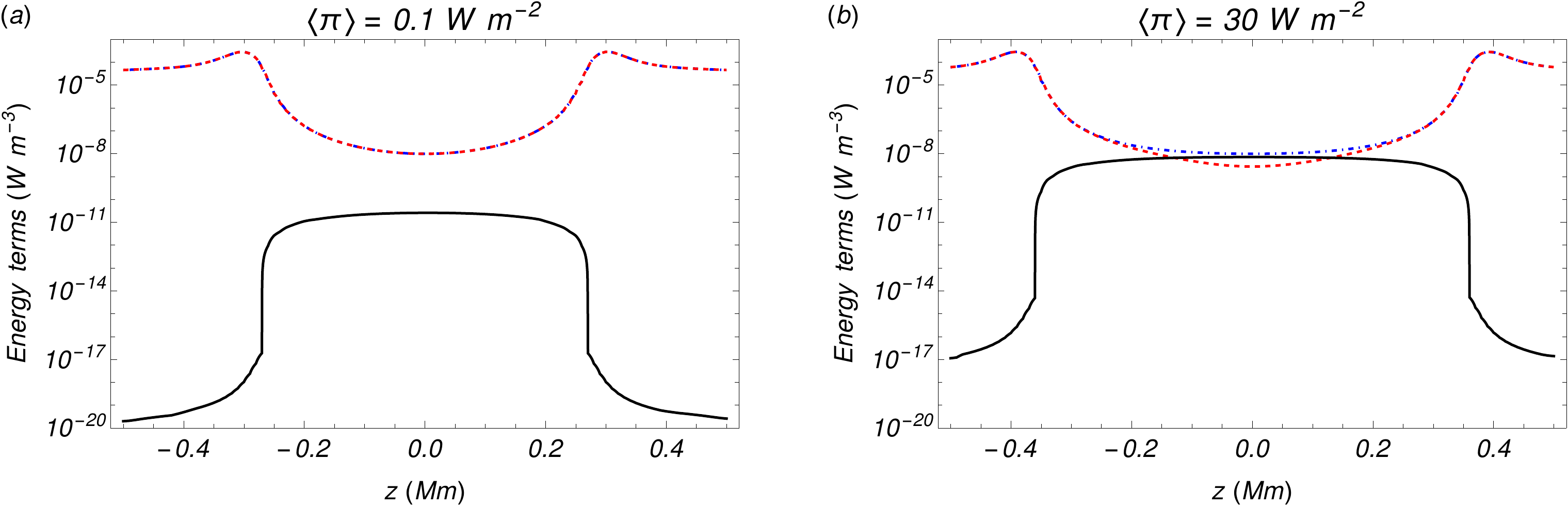}
    \caption{Comparison of the various terms in the energy balance equation (Equation~(\ref{eq:balance})) in models with $T_0=$~9,000~K and two values of the injected wave energy flux: (\textit{a}) $\langle \pi \rangle =$~0.1~W~m$^{-2}$ and (\textit{b}) $\langle \pi \rangle =$~30~W~m$^{-2}$. The solid black line corresponds to the wave heating rate, the dot-dashed blue line denotes radiative losses, and the dashed red line represents the thermal conduction term. Only  a close-up view of the central part and the PCTR of the thread is displayed.}
    \label{fig:energy}
\end{figure*}

Here we explore the role of wave heating in the energy balance. We consider an illustrative example of self-consistent thread models obtained for a central temperature of 9,000~K  and two different injected wave energy fluxes, namely $\langle \pi \rangle =$~0.1 and 30~W~m$^{-2}$. Figure~\ref{fig:energy} compares the importance along the thread of the various terms that appear in the energy balance equation (Equation~(\ref{eq:balance})), namely radiation losses, thermal conduction, and wave heating. To  better visualise the results, Figure~\ref{fig:energy} only displays the cool central part, the PCTR, and the beginning of the coronal part. 

The shape of the wave heating rate is similar for both values of $\langle \pi \rangle$.  The maximum of $\langle Q \rangle$ is located at the centre of the thread, then there is a sharp decrease of several orders of magnitudes in the PCTR, so that the values of $\langle Q \rangle$ in the coronal part of the model are much smaller than in the central cool part. Indeed, as already pointed out, $\langle Q \rangle$ seems to mimic the profile of the Cowling diffusion coefficient, $\eta_{\rm C}$, in the central and PCTR regions, while in the coronal zone it approximately follows  the profile of the Ohm's diffusion coefficient, $\eta$. This highlights the dominance of each mechanism in the different regions.

In the case with $\langle \pi \rangle =$~0.1~W~m$^{-2}$, the wave heating rate is much less important than radiative losses and thermal conduction, even in the central cool part. In this case, wave heating has an almost negligible influence on the energy balance, since radiative losses and thermal conduction essentially compensate each other everywhere along the field line. Radiative loses have a minimum located at the centre of the thread, while their maximum is located inside the PCTR around $T\approx$~58,000 K. In this case, the thermal conduction term just follows the spatial dependence of radiative losses. Comparing the results with the work of \citet{terradas2021thread}, the shapes of the radiative losses and thermal conduction terms obtained here are identical to those obtained in the previous work, although \citet{terradas2021thread} used a constant heating  instead of a function depending on  position.

The situation changes when we consider a larger wave energy flux of $\langle \pi \rangle =$~30~W~m$^{-2}$. Now, a significantly larger heating rate is obtained in the central region, which approaches the minimum of radiative losses. For this large value of the wave energy flux, we are close to the upper boundary of  $\langle Q \rangle$ imposed by Equation~(\ref{eq:second}) to obtain a self-consistent model. Around the thread centre, wave heating compensates a large fraction of radiation losses, while thermal conduction has now a  lower weight on the energy balance. The relevance of wave heating only applies in the central part of the thread, where the plasma is cool and partially ionised, while in the PCTR and, specially, in the coronal region its contribution remains almost negligible and thermal conduction entirely compensates  cooling. These results  are consistent with the work of \citet{melis2021thread}, where the heating rate in the cool central region was estimated to be as important as radiative loses for large enough wave energy fluxes.


As already discussed in Section~\ref{sec:flux}, another  difference between the results for the two considered wave energy fluxes is that the length of the thread increases when $\langle \pi \rangle =$~30~W~m$^{-2}$ compared to the case with $\langle \pi \rangle =$~0.1~W~m$^{-2}$. The computed lengths are $a\approx$~0.59~Mm for $\langle \pi \rangle =$~0.1~W~m$^{-2}$ and $a\approx$~0.77~Mm for $\langle \pi \rangle =$~30~W~m$^{-2}$.

\subsection{Thread length: parameter survey}
\label{sec:param}

In view that the main effect of wave heating is to modify the thread length, here we perform a parameter study on the variation of this quantity as a function of $T_0$ and $\langle \pi \rangle$. Although some results have already been obtained in previous sections, now we carry out a more detailed  investigation.

\begin{figure}[!htbp]
    \resizebox{\hsize}{!}{\includegraphics{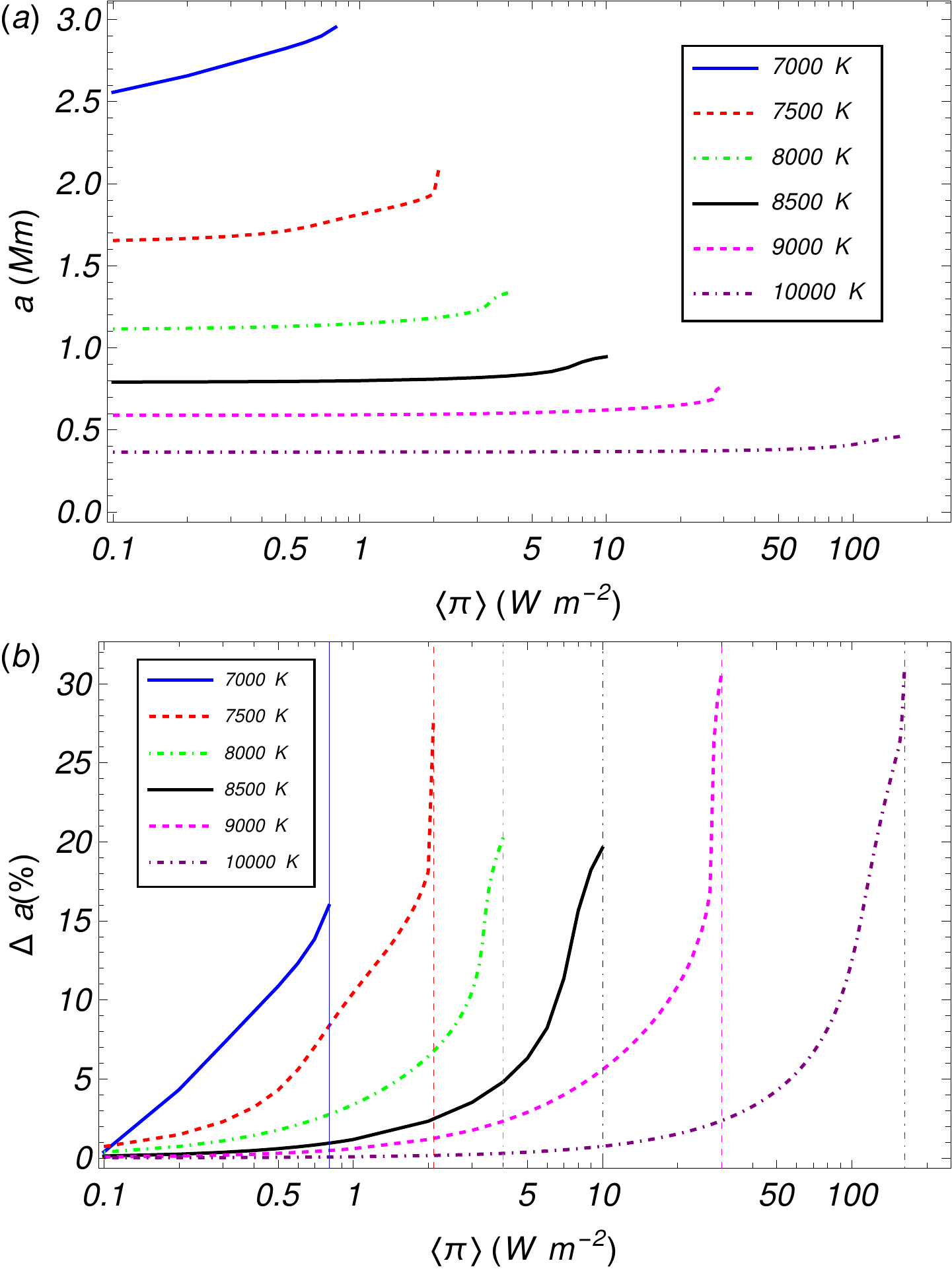}}
    \caption{(\textit{a}) Thread length, $a$, as a function of the injected wave energy flux, $\langle \pi \rangle$, for different values of central temperature.  (\textit{b}) Percentage increase of the thread length with respect to the value without wave heating as a function of  $\langle \pi \rangle$ for the same central temperatures. The vertical lines in panel (\textit{b}) denote the value of $\langle \pi \rangle_{\rm max.}$ corresponding to each temperature.}
    \label{fig:width_f}
\end{figure}

Figure~\ref{fig:width_f}a shows the variation of thread length as a function of the injected wave energy flux for the various central temperatures considered. For each central temperature, the wave energy flux has been varied between $\langle \pi \rangle =$~0.1~W~m$^{-2}$ and the maximum flux that allows the requirement of Equation~(\ref{eq:second}) to be satisfied, hereafter denoted by  $\langle \pi \rangle_{\rm max.}$. Comparing the results for different temperatures, we find that for low central temperatures the thread length is larger than for high central temperatures, as already discussed in Section~\ref{sec:temp}, but the evolution of the thread length when $\langle \pi \rangle$ increases is similar for all the considered central temperatures. Three different regimes are found. First, for $\langle \pi \rangle \ll \langle \pi \rangle_{\rm max.}$, the thread length smoothly increases with  $\langle \pi \rangle$ following an approximate linear dependence. Then, when $\langle \pi \rangle$ approaches $\langle \pi \rangle_{\rm max.}$, the thread length  starts to increase more abruptly in an approximately exponential fashion. Finally, for $\langle \pi \rangle \to \langle \pi \rangle_{\rm max.}$, the thread length  asymptotically tends to infinity. Although our iterative strategy allows us to consider large values of $\langle \pi \rangle$  that differ from $\langle \pi \rangle_{\rm max.}$  by a tiny percentage, the method becomes unstable when $\langle \pi \rangle$ is too close to $\langle \pi \rangle_{\rm max.}$. For this reason, the third regime, i.e., the asymptotic behaviour of the thread length when $\langle \pi \rangle \to \langle \pi \rangle_{\rm max.}$ is only partially captured by the results shown in Figure~\ref{fig:width_f}a. Such asymptotic behaviour can be better visualised in Figure \ref{fig:width_f}b, which displays the percentage increase of the thread length as a function of $\langle \pi \rangle$ with respect to the length obtained in the absence of wave heating. Depending on the central temperature, a percentage increase of the thread length between 15\% and 30\% is obtained for wave energy fluxes just below $\langle \pi \rangle_{\rm max.}$.

\begin{figure}[!htbp]
    \resizebox{\hsize}{!}{\includegraphics{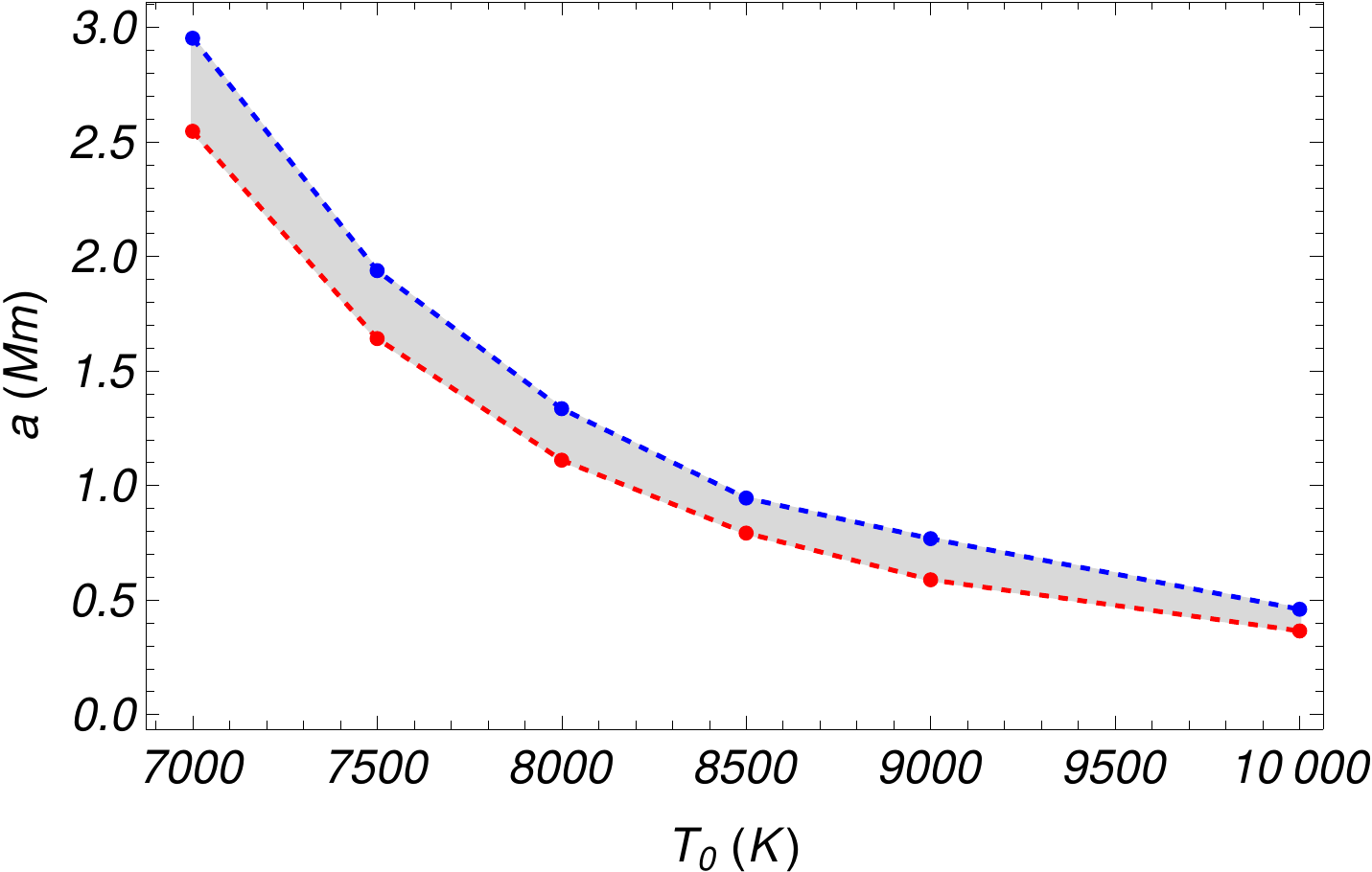}}
    \caption{Thread length, $a$, as a function of the central temperature, $T_0$, in the absence of wave heating (lower red line) and for $\langle \pi \rangle \approx \langle \pi \rangle_{\rm max.}$ (upper blue line). The grey area indicates the range of possible thread lengths for which a self-consistent model can be found.}
    \label{fig:width_t}
\end{figure}

Figure~\ref{fig:width_t} presents the same results of Figure~\ref{fig:width_f} but in a complementary way. Figure~\ref{fig:width_t} displays the thread length as a function of the central temperature in two cases: the length obtained when no wave heating is present, i.e., for $\langle \pi \rangle = 0$, and the length obtained for a wave energy flux just below $\langle \pi \rangle_{\rm max.}$, i.e., at the beginning of the asymptotic regime. As stated before, the thread length decreases when the central temperature increases, which is consistent with the previous results of \citet{terradas2021thread}. For a given central temperature, Figure~\ref{fig:width_t} shows the range of values of the thread length for which our iterative method converges and a self-consistent model is obtained. Threads with smaller lengths are not possible.  Threads with larger lengths are indeed possible, but only in the asymptotic regime  when $\langle \pi \rangle \to \langle \pi \rangle_{\rm max.}$, which happens in a extremely narrow range of values of $\langle \pi \rangle$ just below $\langle \pi \rangle_{\rm max.}$.

\begin{figure}[!htbp]
    \resizebox{\hsize}{!}{\includegraphics{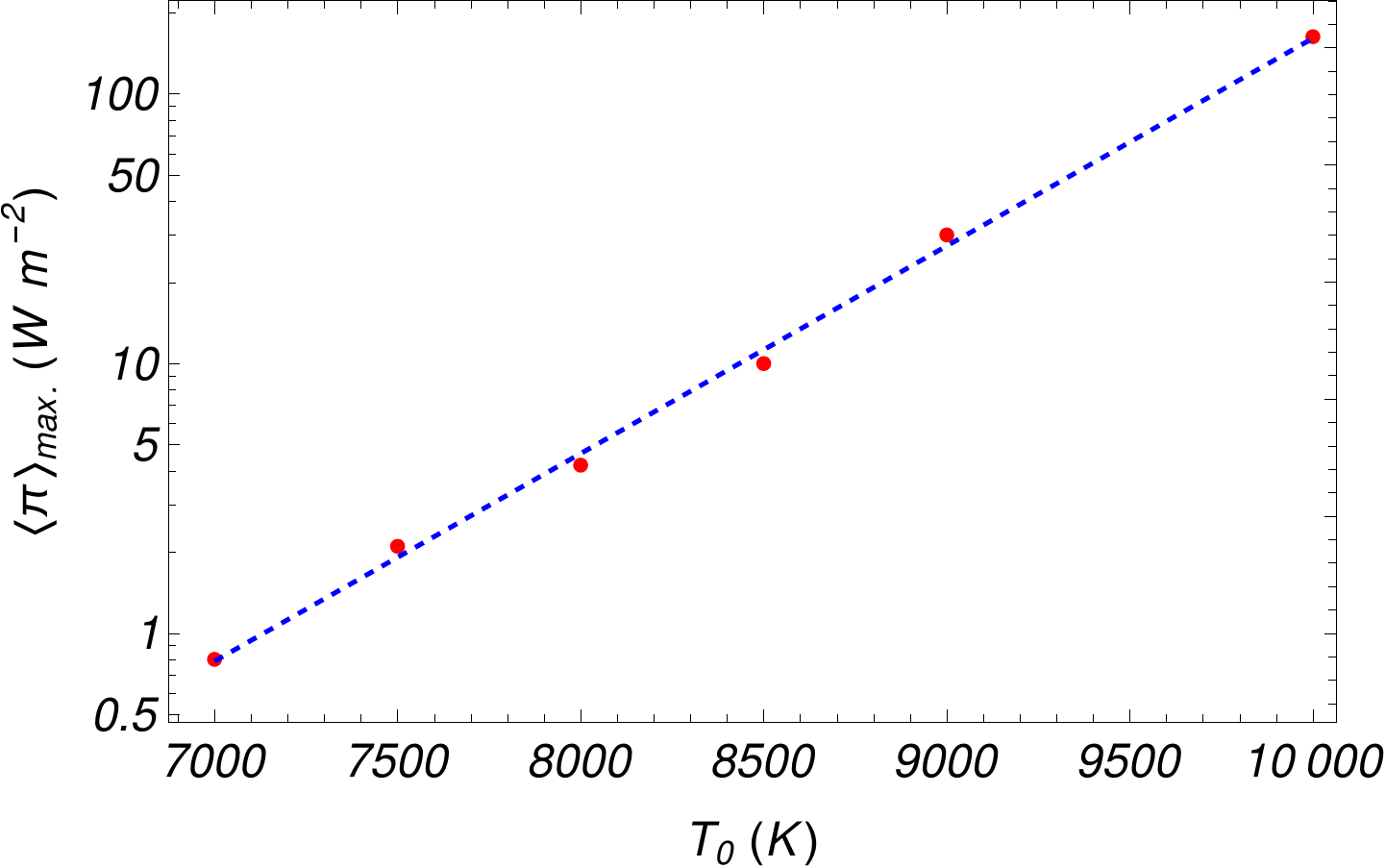}}
    \caption{Maximum wave energy flux injected at the photosphere, $\langle \pi \rangle_{\rm max.}$, as a function of the thread central temperature, $T_0$. The symbols are the results of the computations, while the dashed line in the fit of Equation~(\ref{eq:fit}). Note that the vertical axis is in logarithmic scale.}
    \label{fig:heat_f}
\end{figure}

Another interesting result is that $\langle \pi \rangle_{\rm max.}$ is a function of the central temperature (see Figure~\ref{fig:heat_f}). High central temperatures are able to accommodate larger wave energy fluxes than low central temperatures. Again, the reason for this result can be found in the requirement of Equation~(\ref{eq:second}) that an equilibrium model must satisfy. At $z=0$, the radiative losses provided by Athay's function (Equation~\ref{eq:athay}) increase with $T_0$. Therefore, the maximum wave  heating rate  at $z=0$ that is consistent with Equation~(\ref{eq:second})  also becomes larger when $T_0$ increases. Obviously, larger wave heating rates, $\langle Q \rangle$, are associated with larger wave energy fluxes, $\langle \pi \rangle$. The results of Figure~\ref{fig:heat_f} suggest a power-law dependence of $\langle \pi \rangle_{\rm max.}$ with $T_0$. Therefore, we have performed a least squares fit to the results of Figure~\ref{fig:heat_f} considering the following form,
 \begin{equation}
     {\rm log}_{10} \langle \pi \rangle_{\rm max.} = a T_{0} + b,
     \label{eq:fit}
 \end{equation}
where $a=(770 \pm 18) \times 10^{-6}$ and $b=-5.49 \pm 0.15$ with $\langle \pi \rangle_{\rm max.}$ expressed in W~m$^{-2}$ and $T_0$ in K. The $R^2$ coefficient of the fit is $R^{2} = 0.997$. The fitted function has been over-plotted in Figure~\ref{fig:heat_f}, showing a very good approximation to the numerical data. Hence, for a given central temperature, Equation~(\ref{eq:fit}) can be used to predict the maximum wave energy flux that allows an equilibrium model to exist.

\section{Conclusions}
\label{sec:conclusions}
In this paper we studied 1D models of prominence thin threads that satisfy energy balance including heating due to Alfv\'en waves. The present investigation has been built upon the previous works of \citet{terradas2021thread} and \citet{melis2021thread}, where equilibrium models  and Alfv\'en wave heating were studied separately. The computation of the thread models has been done using an iterative method involving the solution of the energy balance equation together with the Alfv\'en wave equation until convergence to a self-consistent model of a thin thread is achieved. The Alfv\'en waves are assumed to be launched from the photosphere by a broadband driver that injects a prescribed value of the wave energy flux. The waves are then dissipated in the prominence by Ohm's and ambipolar diffusion.

The computed thread models are composed of a cold central region that corresponds to the prominence core (the prominence thread itself), a thin PCTR in which there is a sharp increase of temperature, and an external region with solar coronal properties. The wave heating rate is maximum at the central part, where the plasma is partially ionised and ambipolar diffusion is responsible for wave dissipation. In the outermost coronal part, wave heating is negligible because of the irrelevance of Ohm's diffusion for fully ionised coronal conditions. 

The magnitude of the wave heating rate in the cold central region depends on the wave energy flux injected by the driver at the photosphere, so that the larger the injected energy flux, the larger the heating rate. Increasing the value of the injected energy flux results in obtaining models with longer cold regions, i.e., longer threads. Self-consistent models can be obtained until the injected wave energy flux produces a heating rate at the thread centre that becomes equal to radiative losses. We performed a parametric survey in order to explore how the thread length correlates with the injected wave energy flux and the central temperature. The thread length decreases with increasing central temperature and increases with the injected wave energy flux. The maximum value of the wave energy flux that allows self-consistent models to exist has been found to increase with the value of the central temperature, so that hot threads can accommodate more wave heating than cool ones.

We have used the injected wave energy flux as a free parameter of the model. Unfortunately, there are no determinations of the Alfv\'en wave energy flux driven at the footpoints of the magnetic structure of prominences. In intense photospheric flux tubes, calculations of the transverse wave energy flux driven by horizontal motions show that the flux generated within the photospheric flux tubes can be as large as $\sim 10^6$~W~m$^{-2}$ \citep[see, e.g.,][]{spruit1981,Ulmschneider2000,noble2003,Shelyag2011}. When averaged over the entire photosphere and considering the photospheric filling factor, the resulting averaged driven flux is of the order of $\sim 10^4$~W~m$^{-2}$ \citep[see, e.g.,][]{depontieu2001,goodman2011,tu2013heat,arber2016chromo}. This estimated wave energy flux applies in the case of photospheric bright points, but it is probably much larger than the wave energy flux driven in the case of prominences. If this large energy flux were applicable to prominences, then our results indicate that no equilibrium models of threads would be possible: wave heating would be simply too large. Stable threads could not sustain such large wave energy fluxes. In essence, the existence or absence of equilibrium models depends upon the balance between wave heating and radiation losses at the thread centre. While radiation losses mainly depend on the local properties of the thread, the magnitude of wave heating is determined by how the waves are driven at the photospheric regions where the prominence magnetic field is anchored.  Recently, \citet{li2022} reported Alfv\'enic waves along prominence threads with an estimated energy flux of 16.2--37.3~W~m$^{-2}$, which is more compatible with the wave energy fluxes needed to obtain equilibrium models.

For simplicity and to consider the same set-up as in \citet{melis2021thread}, a perfectly reflecting boundary has been implemented at the right end of the thread instead of considering the presence of two different drivers located at both ends. The expected results of adding  a second driver to the model are straightforward.  If the second driver injects the same energy flux as the first one, the heating rate in the cool region would  double its value. In turn, the maximum value of the wave energy flux to obtain a self-consistent model would then need to account for the sum of the fluxes injected by both drivers.

A limitation of this work is that the computation of the wave heating rate assumes the stationary state of wave propagation. In other words, it has been assumed that the waves have been propagating and reflecting along the thread for enough time to reach a stationary pattern. An alternative approach would be to solve the time-dependent problem, for which we should resort to full numerical simulations.  The comparison of time-dependent results with those of the  stationary problem is relevant and may be tackled in the future.

Although the present method of constructing thread models heated by Alfv\'en waves provides interesting results, the background configuration is still simple compared to the observed complexity of prominences. A further extension would be to consider 2D models in cylindrical geometry, which should include the perpendicular component of thermal conduction. Concerning the Alfv\'en waves, the use of 2D models would introduce the presence of resonances in the Alfv\'en continuum. Adding these new ingredients  would allow us to obtain more accurate information about the efficiency of wave heating in solar prominences and the conditions for the existence of equilibrium models. This could also be undertaken in a follow-up  work.

Finally, we connect with the comment made in the Introduction about the possible role that Alfv\'en wave heating may have to complement radiative heating in prominences. Here we have computed thread models with realistic lengths and central temperatures without including radiative heating. This suggests that Alfv\'en wave heating may play not only a complementary role but an important part in the energy balance of prominence threads that future works should keep investigating. Ideally, future models should include both radiative and wave heating, which is a challenging task.

\begin{acknowledgement}
This publication is part of the R+D+i project PID2020-
112791GB-I00, financed by MCIN/AEI/10.13039/501100011033. This research was supported by the International Space Science Institute (ISSI) in Bern, through ISSI International Team project \#457 (The Role of Partial Ionization in the Formation, Dynamics and Stability of Solar Prominences). We thank Ram\'on Oliver for some comments regarding the importance of correctly resolving the PCTR. We acknowledge the unknown referee for very constructive comments that helped us to improve the paper.
\end{acknowledgement}

\bibpunct{(}{)}{;}{a}{}{,} 
\bibliographystyle{aa} 
\bibliography{refs} 
\end{document}